\documentclass[12pt]{article}
\usepackage{amsmath,amssymb,amsfonts}
\usepackage{color,graphicx,cite,soul}
\usepackage{array,booktabs,longtable}
\usepackage{caption,microtype,url}

\graphicspath{{figs/}}

\evensidemargin \oddsidemargin
\allowdisplaybreaks
\usepackage{graphicx}
\usepackage{epsfig}
\usepackage{dcolumn}
\usepackage{bm}
\usepackage{here}
\usepackage{braket}
\usepackage{amsmath}
\def\beq{\begin{equation}}
\def\eeq{\end{equation}}
\def\bea{\begin{eqnarray}}
\def\eea{\end{eqnarray}}
\def\ba{\begin{array}}
	\def\ea{\end{array}}
\def\bi{\begin{itemize}}
	\def\ei{\end{itemize}}
\def\be{\begin{enumerate}}
	\def\ee{\end{enumerate}}
\def\bc{\begin{center}}
	\def\ec{\end{center}}
\def\bt{\begin{table}}
	\def\et{\end{table}}
\def\btb{\begin{tabular}}
	\def\etb{\end{tabular}}

\begin{document}

\def\thefootnote{\fnsymbol{footnote}}
\thispagestyle{empty}

\begin{flushright}
	 RBI-ThPhys-2021-23 \\
	 CERN-TH-2021-080
	 \end{flushright}

\begin{center}

{\large\sc 
	{\bf Di-Higgs production \boldmath{($\gamma \gamma \to h h$)} in Composite Models}
		\footnote{Talk presented at the International Workshop on Future Linear Colliders (LCWS2021), 15-18 March 2021. C21-03-15.1.}
}	
	\vspace{1.5 cm}
	
{\sc
A.~Bharucha$^1$
	\footnote{email: aoife.bharucha@cpt.univ-mrs.fr},
G.~Cacciapaglia$^2$ 
	\footnote{email: g.cacciapaglia@ipnl.in2p3.fr},
A.~Deandrea$^2$
	\footnote{email: deandrea@ipnl.in2p3.fr},
N.~Gaur$^3$
	\footnote{email: naveengaur@dsc.du.ac.in}, \\ 
D.~Harada$^4$
	\footnote{email: dharada@irb.hr}
	\footnote{speaker}, 
F.~Mahmoudi$^{2,5}$
	\footnote{email: nazila@cern.ch },
K.~Sridhar$^{6,7}$
	\footnote{email: sridhar@theory.tifr.res.in}
}

	\vspace{1cm} 
	
{\sl 
  $^1$ Aix Marseille Univ, CNRS, CPT, Marseille, France, 13288 Marseille, France \\
  \vspace*{0.1cm}
  $^2$ Universit\'e de Lyon, Universit\'e Claude Bernard Lyon 1, CNRS/IN2P3, \\
  Institut de Physique des 2 Infinis de Lyon, UMR 5822, F-69622, Villeurbanne, France \\
  \vspace*{0.1cm}
  $^3$ Department of Physics, Dyal Singh College (University of Delhi), Lodhi Road, \\ 
  New Delhi, 110003, India \\
  \vspace*{0.1cm}
  $^4$ Rudjer Boskovic Institute, Division of Theoretical Physics, \\ 
   Bijenicka cesta 54, 10000 Zagreb, Croatia \\
  \vspace*{0.1cm}
  $^5$ Theoretical Physics Department, CERN, CH-1211 Geneva 23, Switzerland \\
   \vspace*{0.1cm}
   $^6$ Department of Theoretical Physics, Tata Institute of Fundamental Research, \\
   Homi Bhabha Road, Colaba, Mumbai 400005, India
    \\ 
   $^7$ KREA University, Sri City, Andhra Pradesh-517646, India
}

\end{center} 



\vspace*{0.1cm}

\begin{abstract}
\noindent
In Standard Model (SM) Higgs Boson pair production initiated by photons ($\gamma \gamma \to h h$) is loop-generated process
and thereby very sensitive to any new couplings and particles that may come in loops. The Composite Higgs Models 
provide an alternate mechanism to address the hierarchy problem of SM where Higgs instead of being an elementary field 
could be a bound state of a strongly interacting sector. These set of models apart from modifying the SM Higgs 
couplings could also introduce new effective couplings that can have substantial impact on the loop processes. 
In this work we have studied the impact of such modifications by Composite Higgs models in $\gamma\gamma \to h h$ production process. 
\end{abstract}

\def\thefootnote{\arabic{footnote}}
\setcounter{page}{0}
\setcounter{footnote}{0}

\newpage

\section{Introduction}\label{sect1}

After the discovery of the Standard Model (SM) like Higgs boson at the CERN Large Hadron Collider (LHC)~\cite{Aad:2012tfa,Chatrchyan:2012ufa,Aad:2015zhl}, the question of the origin of the scalar sector has become more crucial in understanding the physics Beyond the Standard Model (BSM).
Model building efforts and specific searches aim at discovering features that can shed light on the fundamental mechanism behind the Higgs sector.
One of the popular extension of the SM consists of replacing the Higgs sector by a new strong-interaction at the electroweak (EW) scale, giving rise to EW symmetry breaking of  dynamical origin~\cite{Weinberg:1975gm,Dimopoulos:1979es,Eichten:1979ah}.
In this framework, the relative lightness of the Higgs boson is ensured by the pseudo Nambu-Goldstone Boson (pNGB) nature of this particle, which stems from the broken global symmetry~\cite{Kaplan:1983fs}.
In recent years detailed models were proposed based on a holographic description~\cite{Contino:2003ve,Agashe:2004rs,Agashe:2005dk,Contino:2006qr}, or based on an underlying gauge-fermion theory, where the global symmetry is broken by the bilinear condensate of techni-fermions~\cite{Peskin:1980gc,Ryttov:2008xe,Galloway:2010bp,Cacciapaglia:2014uja}.
The underlying theories are designed to feature a vacuum alignment that does not break the SM gauge symmetry and a Higgs doublet in the pNGB sector (for a review see~\cite{Cacciapaglia:2020kgq}), contrary to the old-school Technicolor theories that break the electroweak symmetry at the condensation scale without a Higgs boson~\cite{Weinberg:1975gm}.
The exploration of these BSM scenarios is an active research subject at present and in the future colliders.

The BSM searches at the LHC have not yet led to a discovery of the new particles or a new phenomena, a sign that BSM physics is subtler and/or fainter than what was originally expected.
The electron-positron collider option provides a rather clean experimental conditions in comparison to the LHC, where the quantum chromodynamics (QCD) background is intense and hard to master.
There are proposals for the future electron-positron colliders both circular designs, as for example, the FCC-ee at CERN~\cite{Abada:2019lih,Abada:2019zxq} and the CEPC in China~\cite{CEPCStudyGroup:2018rmc,CEPCStudyGroup:2018ghi} and linear ones, such as the International Linear Collider (ILC)~\cite{Baer:2013cma}.
In addition to the electron-positron colliders, the Photon Linear Collider (PLC) is also considered as an optional experiment of the ILC.
Compton back-scattering of laser photons on electrons at the ILC allows the production of high energy photons. These photon beams can reach energies close to those of the initial electrons. The PLC is therefore a compelling option for the ILC~\cite{Ginzburg:2019yws}.
The possibility of measuring the triple Higgs coupling via the $\gamma\gamma\to hh$ process at the PLC has been widely discussed in the literature, as the sensitivity of this channel to the Higgs self-coupling is maximal at the threshold $2m_{h}$. This process sensitivities are greater than the sensitivities 
achieved at the electron positron direct collision in processes such as $e^{+}e^{-}\to Zhh$ and $e^{+}e^{-}\to\nu\bar{\nu}hh$ in a wide range of center of mass energies~\cite{Jikia:1992zw,Belusevic:2004pz}.
The $\gamma\gamma\to hh$ process~\cite{Jikia:1992mt} is induced by top quark and $W^{\pm}$ boson loops, details are given in Ref.~\cite{Asakawa:2008se,Asakawa:2010xj,Kawada:2012uy}.
In the SM, these loops have destructive interference.
In BSM scenarios, there can be a substantial change in the production cross-sections due to the change in the cancellations between top quark and 
$W$ boson loops ~\cite{Asakawa:2008se,Asakawa:2010xj,Kawada:2012uy}.
Compared to other di-Higgs channels at tree level, the photon fusion channel is of more interest because it provides access to more couplings of the Higgs which enter in the loops~\cite{Asakawa:2010xj}.
This is particularly attractive in BSM scenarios, where the top quark plays an important role due to its large mass.

The $\gamma\gamma\to hh$ process at the PLC could also be an  excellent probe of Composite Higgs Models (CHMs).
In fact, all models in this class feature modifications of the Higgs couplings to gauge bosons and top quark, which can alter the 
cancellations that occur in the SM.
%
%
Composite resonances, if relatively light, can also provide additional contributions.
This in turn has effects both on the cross-section and on the helicity distributions, allowing models based on the composite Higgs idea to be further constrained (or discovered).
%
%
%

In this paper, we study the impact of CHMs on the di-Higgs production process via photon fusion ($\gamma\gamma\to hh$). The paper is organized
as follows: 
In Sec.~\ref{section:2}, we have presented numerical results for the cross-section of the di-Higgs production in CHMs at the PLC.
In Sec.~\ref{section:3}, the results of statistical sensitivity of the process are presented.  We finally conclude in Sec.~\ref{section:4}.

\section{Di-Higgs production in Composite Models at the PLC}\label{section:2}

The Feynman diagrams for this process in the CHMs are shown in Ref.~\cite{Bharucha:2020bhy}.
The differential cross section for the di-Higgs process $\gamma\gamma\to hh$ can be written in terms of the helicity amplitudes as
\begin{equation}
\frac{d\hat{\sigma}(\lambda_1,\lambda_2)}{d\hat{t}} = \frac{\alpha^{2}\alpha_{W}^{2}}{32 \pi\hat{s}^{2}} \left| {\cal M}(\lambda_1,\lambda_2) \right|^{2} \,.
\label{sec2:eq_1} 
\end{equation}
In the CHMs, the helicity amplitudes ${\cal M}(\lambda_1,\lambda_2)$ for the initial photon helicities $\lambda_{1}$ and $\lambda_{2}$ ($\lambda_{i}=+$ or $-$) are given in Ref.~\cite{Bharucha:2020bhy}.
The total cross section of $e^{+}e^{-}\to\gamma\gamma\to hh$ is calculated by convoluting the photon luminosity function
\begin{equation}
\sigma = \int_{4m_{h}^{2}/s}^{y_{m}^{2}} d\tau \frac{dL_{\gamma\gamma}}{d\tau} \left[ \frac{1+\xi_{1}^{\gamma}\xi_{2}^{\gamma}}{2} \hat{\sigma}_{++}(\hat{s}) + \frac{1-\xi_{1}^{\gamma}\xi_{2}^{\gamma}}{2} \hat{\sigma}_{+-}(\hat{s}) \right] \,,
\end{equation}
where $\xi^{\gamma}_{(1,2)}$ are the mean photon helicities, and the differential luminosity takes the form
\begin{eqnarray}
\frac{dL_{\gamma\gamma}}{d\tau} =  \int_{\tau/y_{m}}^{y_{m}} \frac{dy}{y} f_{\gamma}(x,y) f_{\gamma}(x,\tau/y) \,,
\end{eqnarray}
where $\tau=\hat{s}/s$, $y=E_{\gamma}/E_{b}$ with $E_{\gamma}$ and $E_{b}$ being the energy of photon and electron beams respectively, and the maximal energy fraction of photon $y_{m}=x/(1+x)$ with $x=4E_{b}\omega_{0}/m_{e}^{2}$ where $\omega_{0}$ is the laser photon energy and $m_{e}$ is the electron mass. 
The photon luminosity spectrum $f_{\gamma}(x,y)$ is given by \cite{Ginzburg:1982yr}.
In our analysis, we set the dimensionless parameter $x=4.8$ ($y_{m}=0.82$) and a very high degree of electron beam polarization (90\%), i.e. $\lambda_{e_{1}}=\lambda_{e_{2}}=0.45$, $\lambda_{\gamma_{1}}=\lambda_{\gamma_{2}}=-1$~\cite{Jikia:1992mt}.
%

\subsection{Minimal Composite Higgs Models}\label{section:2-1}

\begin{table}[tb]
	\begin{center} 	\begin{tabular}{|c|c|c|c|c|c|} \hline 
			Model & $h f \bar{f} (c_f)$ & $h h f \bar{f} (c_{2f})$ & $ h W^+ W^- (c_v)$ & $h h W^+ W^- (c_{2v})$
			          &  $c_{3h}$ \\ \hline 
			MCHM4 \cite{Agashe:2004rs} & $\sqrt{1 - \xi}$  &  $ -\xi$  & $\sqrt{1-\xi}$ & $1-2\xi$ & $\sqrt{1-\xi}$    \\ \hline 
			MCHM5 \cite{Contino:2006qr} & $\frac{1 - 2\xi}{\sqrt{1 - \xi}}$  &  $-4 \xi$  & $\sqrt{1-\xi}$ & $1-2\xi$ 
			         & $\frac{1 - 2 \xi}{\sqrt{1 - \xi}}$ \\ \hline      
					\end{tabular}
	\caption{The Higgs couplings as a function of $\xi = v^2/f^2 \equiv \sin^2 \theta$, with $\theta$ being the misalignment angle.	\label{table:1} } 
	\end{center}
\end{table}

\begin{figure}[tb]
	\begin{center}
    $\begin{array}{cc}
    {\bf MCHM4} & {\bf MCHM5} \\
    \hspace*{-1cm}
    \epsfig{file=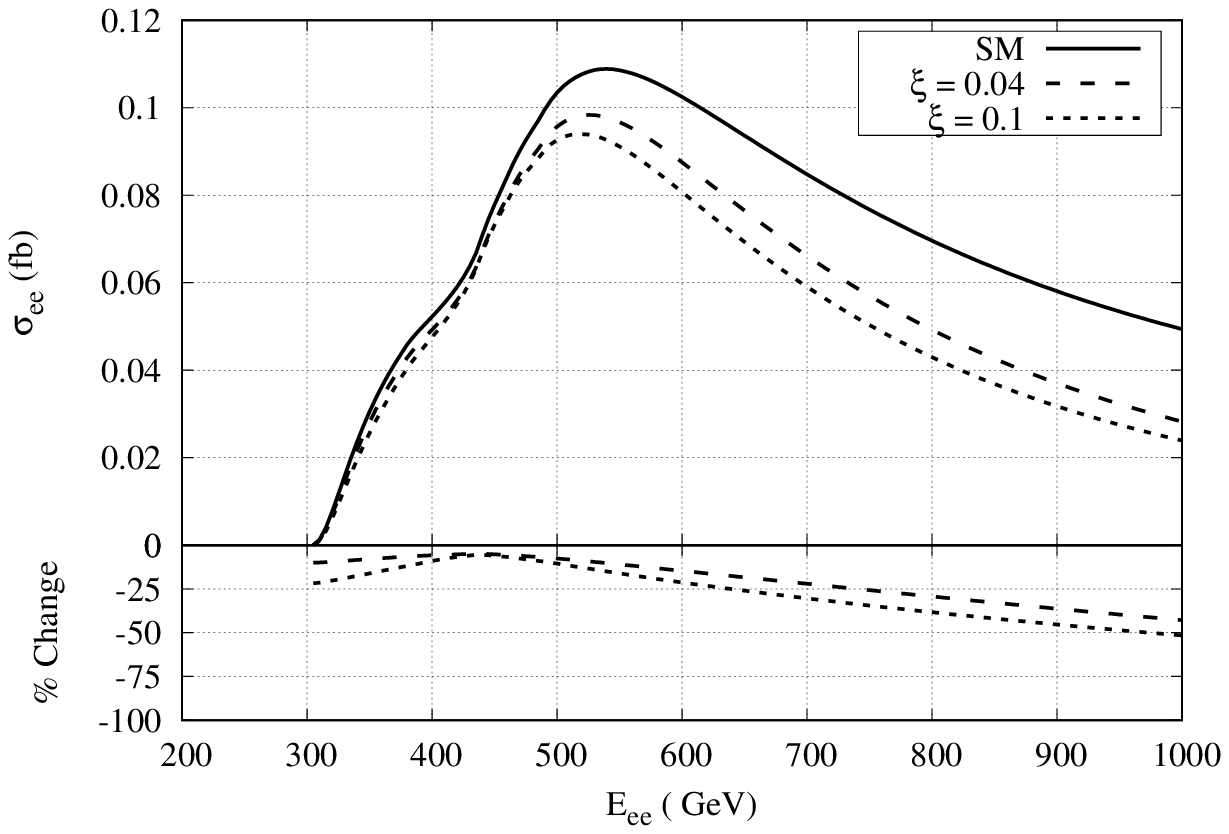,width=0.52\textwidth}  &  
    \epsfig{file=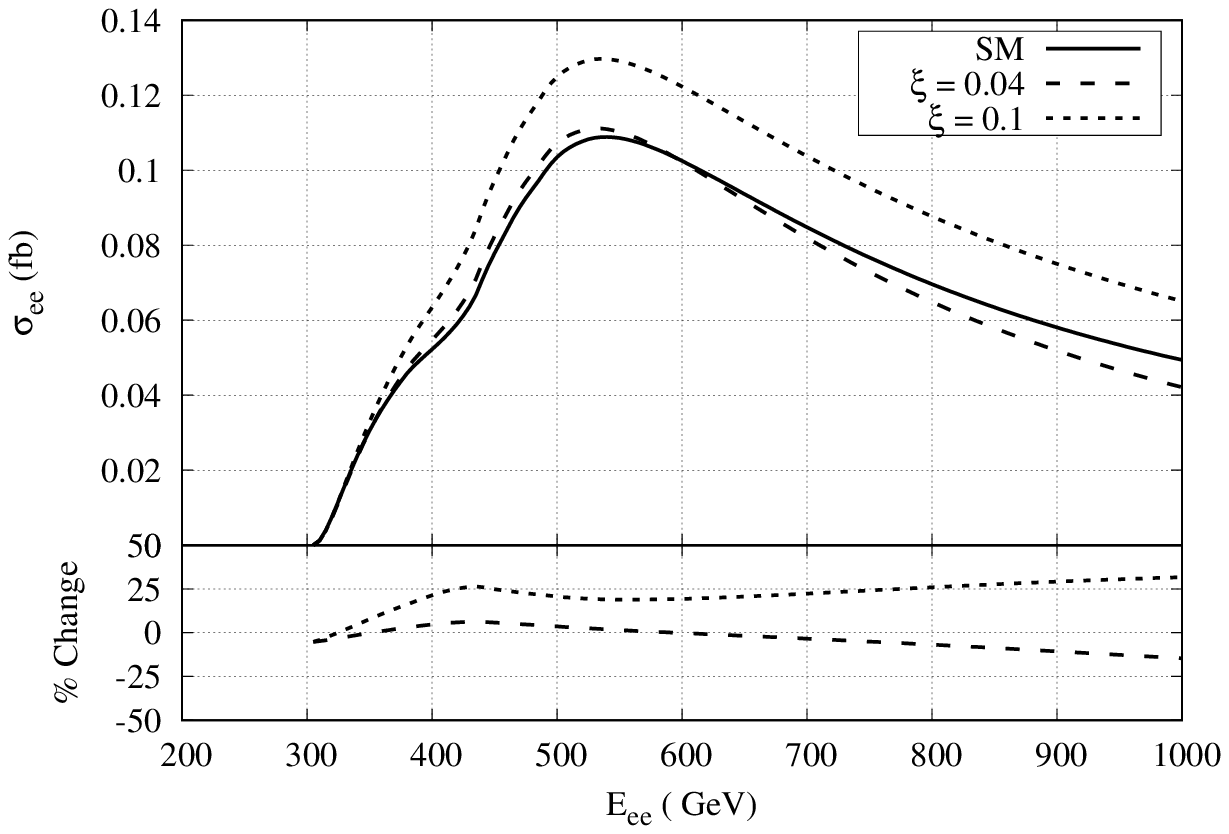,width=0.52\textwidth}	
    \end{array}$
	\caption{$e^+ e^- \to h h$ cross-section as a function of the electron-positron center of mass energy $E_{ee}$ in the MCHM4 (left panel) and MCHM5 (right panel) benchmarks.}  	\label{fig:sec3_2} 
	\end{center}
\end{figure}

In our analysis we have first considered the minimal CHMs, where the vacuum is only misaligned along one direction that 
breaks the EW symmetry. This implies that the composite Higgs does not mix with other pNGBs, and that the misalignment can be described 
in terms of a single angle $\theta$~\cite{Kaplan:1983fs}, defined as follows:
\beq
v = f\sin\theta \,.
\eeq
The above equation provides the relation between the EW symmetry breaking scale $v$ and the compositeness scale $f$.
This definition is independent of the coset ${\cal G}$/${\cal H}$.
The masses of the EW gauge bosons $W^{\pm}$/$Z$ are given by:
\beq
m_{W}^{2}(\theta) = \frac{g^{2}f^{2}}{4}\sin^{2}\theta \equiv \frac{g^{2}v^{2}}{4} \,,\qquad
m_{Z}^{2}(\theta) = \frac{1}{c_{W}^{2}}m_{W}^{2}(\theta) \,,
\eeq
where the relation between the two is ensured by the custodial symmetry embedded in ${\cal G}$/${\cal H}$.
The coset-independence of the relation between the $W^{\pm}$/$Z$ masses and the compositeness scale $f$ leads to universal couplings of the Higgs boson to the EW gauge bosons.
These couplings can be elegantly expressed in terms of derivatives with respect to the misalignment angle $\theta$, as follows:
\bea
g_{WWh} &=& \frac{1}{f} \frac{\partial m_W^2 (\theta)}{\partial \theta} = \frac{2 m_W^2}{v} \cos \theta \,, \\
g_{WWhh} &=& \frac{1}{f^2} \frac{\partial^2 m_W^2 (\theta)}{\partial \theta^2} = \frac{2 m_W^2}{v^2} \cos 2\theta \,,
\eea
For convenience, we will use the parameterisation as given in the Refs.~\cite{Grober:2010yv, Contino:2012xk}, which reads
\beq
{\cal L} = m_W^2 W^+_\mu W^{-,\mu} \left( 1 + 2 c_v \frac{h}{v} + c_{2v}  \frac{h^2}{v^2} + \dots \right)\,,
\label{eq:gauge1} 
\eeq
with
\beq
c_v = \cos \theta = \sqrt{1-\xi}\,, \qquad c_{2v} = \cos 2\theta = 1-2\xi\,;
\eeq
where $\xi = v^2/f^2 \equiv \sin^2 \theta$.

The couplings of the composite Higgs to SM fermions are not universal and can also be expressed in terms of derivatives with 
respect to the misalignment angle $\theta$, as follows:
\bea
g_{ffh} = \frac{1}{f} \frac{\partial m_t (\theta)}{\partial \theta}\,, \quad 
g_{ffhh} = \frac{1}{f^2} \frac{\partial^2 m_t (\theta)}{\partial^2 \theta}\,, \quad \dots
		\label{eq:2} 
\eea 
These expressions depend on the details of the model, and in particular on the dependence of the top mass on the misalignment angle $\theta$.
We will consider two scenarios: the first is realized in the SO(5)/SO(4) CHMs with top partners in the spinorial representation of the 
global SO(5), this model is termed as MCHM4. For this model, we have:
\beq
m_t (\theta) = \frac{\lambda f}{\sqrt{2}} \sin \theta \;\; \Rightarrow \;\; \left\{ \begin{array}{l}
g_{ffh} =\displaystyle \frac{m_t}{v} \cos \theta\,, \\
g_{ffhh} =\displaystyle  - \frac{m_t}{v^2} \sin^2 \theta\,.
\end{array} \right.
\eeq
The second case is realized in the SO(5)/SO(4) model with the fundamental representation of SO(5) (MCHM5), for which:
\beq
m_t (\theta) = \frac{\lambda f}{\sqrt{2}} \sin 2\theta \;\; \Rightarrow \;\; \left\{ \begin{array}{l}
g_{ffh} =\displaystyle  \frac{m_t}{v} \frac{\cos 2\theta}{ \cos \theta}\,, \\
g_{ffhh} =\displaystyle  - \frac{m_t}{v^2} 4 \sin^2 \theta\,.
\end{array} \right.
\eeq
For convenience, we will parameterise the coupling modifier following Ref. \cite{Contino:2010mh} as:
\begin{equation}
{\cal L} =  -m_t \left( \bar{t}_L t_R \right)
\left( 1 + c_f \frac{h}{v} +  \frac{c_{2f}}{2}  \frac{h^2}{v^2} + \dots \right) + h.c. 
\label{eq:fermion1} 
\end{equation}

In Fig.~\ref{fig:sec3_2}, we report the $e^{+}e^{-}$ cross section as a function of the electron-positron center of mass energy for two values of $\xi=0.1$, 0.04 and compared to the SM value.
The main difference between the MCHM4 and MCHM5 benchmarks is a larger top quartic coupling (c.f. Table~\ref{table:1}) for the latter.
In the MCHM4, we observe a systematic decrease in the total cross-section, with sizeable  effects emerging for center-of-mass energies above 500 GeV. Thus, this scenario can only  be tested at a high-energy version of the collider. On the other hand, the MCHM5 can  feature an increase in the cross-section compared to the SM one, driven by the $\bar{t}thh$ coupling $c_{2f}$.
This effect can go up to 20\% above the $\bar{t}t$ threshold for $\xi=0.1$.

\subsection{Heavy scalar}\label{section:2-2}

The presence of a rather light scalar resonance in the spectrum of CHMs has been shown to help in reducing the constraints on the misalignment angle $\theta$~\cite{BuarqueFranzosi:2018eaj}. We have analyzed this scenario where a singlet scalar which is a resonance of the composite sector is present. 
The presence of a relatively light scalar can be encoded in the addition of a second heavier Higgs $H$, with couplings parameterised 
in analogy to those of the SM Higgs:
\begin{multline}
\mathcal{L} \supset m_W^2 W_\mu^+ W^{-,\mu} \left( 1 + 2 c_v \frac{h}{v} + 2 c_v^H \frac{H}{v} +  c_{2v} \frac{h^2}{v^2} + \dots \right)  \\
- m_t \bar{t} t \left( 1 + c_f \frac{h}{v} + c_t^H \frac{H}{v} + \frac{c_{2t}}{2} \frac{h^2}{v^2} +  \dots \right)\,.
\end{multline}
The coefficients $c_{x}$ and $c_{x}^{H}$ ($x=f,v$) can be computed in specific scenarios, and take into account the mixing between the two states.
There is also a derivative coupling of $H$ to two Higgs, which is relevant for di-Higgs pair production, that can be parameterised as:
\beq \label{derHhh}
\mathcal{L} \supset c_{Hhh}\ H \partial_\mu h \partial^\mu h \to - \frac{1}{2} c_{Hhh} \frac{\hat{s} - 2 m_h^2}{v}\ H h h\,.
\eeq
where $\hat{s}$ is the invariant mass of the two $h$ system (i.e., the center of mass energy of the partonic process).

\begin{table}[htb]
	\begin{center}
		\begin{tabular}{|c|c|c|c|c|c|c|} \hline
			Benchmark 1 & \multicolumn{6}{c|}{$m_H = 610$~GeV, $\xi = 0.306$, $\Gamma_H = 498$~GeV, $k'_G = 1.5$ } \\ \hline
			& $c_f/c_f^{H}$ & $c_{2f}$ & $c_v/c_v^{H}$ & $c_{2v}$
			&  $c_{3h}$ & $c_{Hhh}$\\ \hline
			$h$  & $0.9199$  &  $-0.7814$  & $0.8791$ & $0.5562$ & $\lambda_h$  & $-$  \\ \hline
			$H$  & $3.507$  &  $\dots$  & $0.3054$ & $\dots$
			& $-$ & $0.4149$ \\ \hline       \hline
			Benchmark 2 & \multicolumn{6}{c|}{$m_H = 800$~GeV, $\xi = 0.197$, $\Gamma_H = 350$~GeV, $k'_G = 1.8$ } \\ \hline
			& $c_f/c_f^{H}$ & $c_{2f}$ & $c_v/c_v^{H}$ & $c_{2v}$
			&  $c_{3h}$ & $c_{Hhh}$\\ \hline
			$h$  & $0.9102$  &  $-0.4627$  & $0.9305$ & $0.7381$ & $\lambda_h$ & $-$    \\ \hline
			$H$  & $2.368$  &  $\dots$  & $0.3109$ & $\dots$
			& $-$ & $0.4001$\\ \hline   \hline
			Benchmark 3 & \multicolumn{6}{c|}{$m_H = 1000$~GeV, $\xi = 0.0646$, $\Gamma_H = 47.6$~GeV, $k'_G = 1.$ } \\ \hline
			& $c_f/c_f^{H}$ & $c_{2f}$ & $c_v/c_v^{H}$ & $c_{2v}$
			&  $c_{3h}$ & $c_{Hhh}$\\ \hline
			$h$  & $0.9572$  &  $-0.1498$  & $0.9741$ & $0.9038$ & $\lambda_h$ & $-$    \\ \hline
			$H$  & $0.6896$  &  $\dots$  & $0.0511$ & $\dots$
			& $-$ & $0.1270$\\ \hline
		\end{tabular}
	\caption{Couplings of the Higgs $h$ and of the heavier state $H$, for 3 benchmark points. The parameter $k'_G$ characterizes the coupling of the heavy resonance to the gauge bosons (see Ref.~\cite{BuarqueFranzosi:2018eaj} for more details).\label{table:BM12}}
	\end{center}
\end{table}

\begin{figure}[htb]
	\begin{center}
	\hspace*{-1cm}
		\epsfig{file=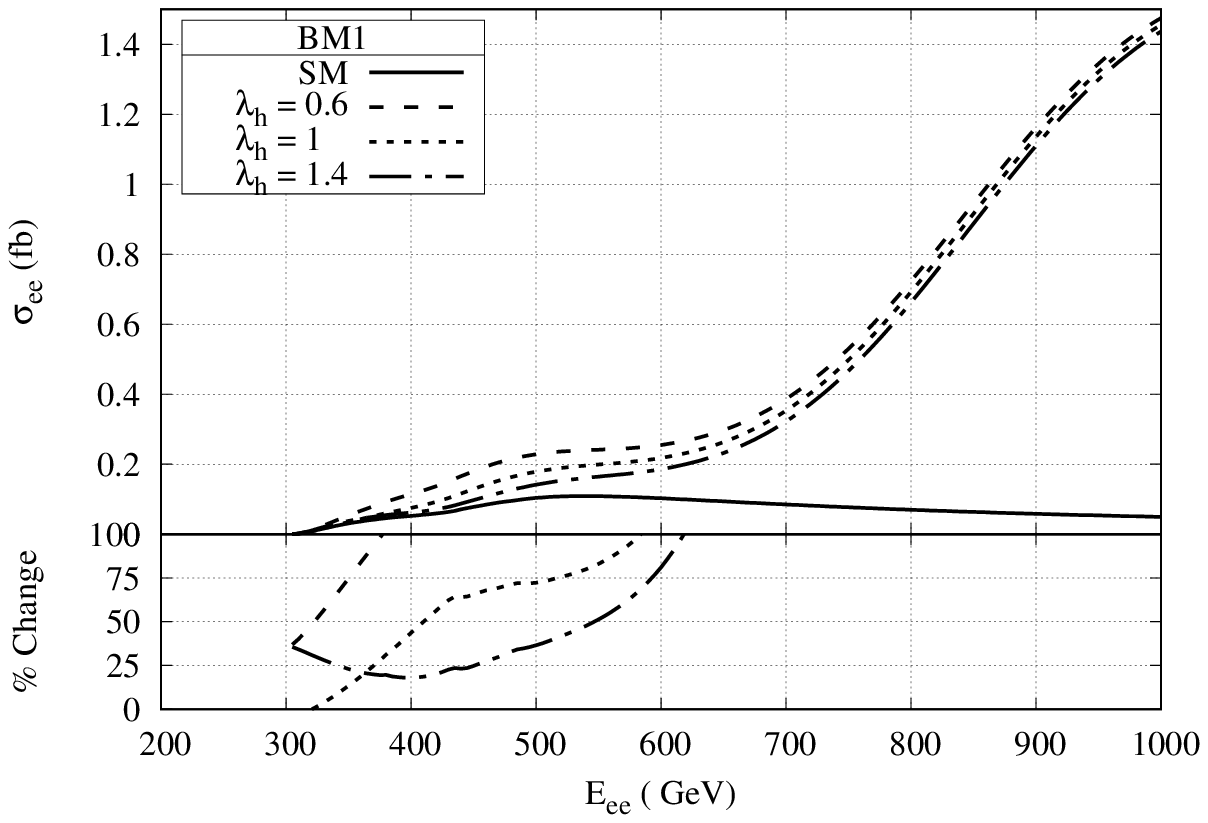,width=0.52\textwidth} 	
		\epsfig{file=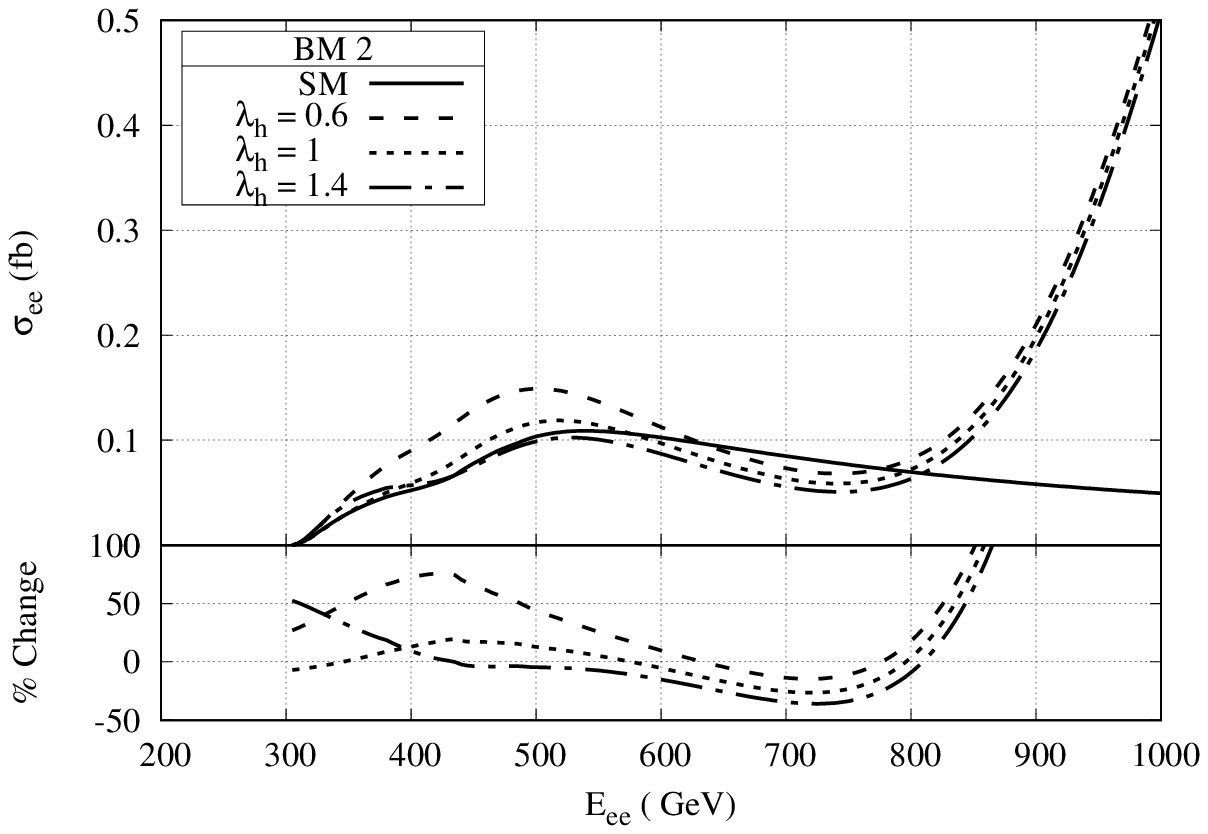,width=0.52\textwidth} 	\\ 
		\epsfig{file=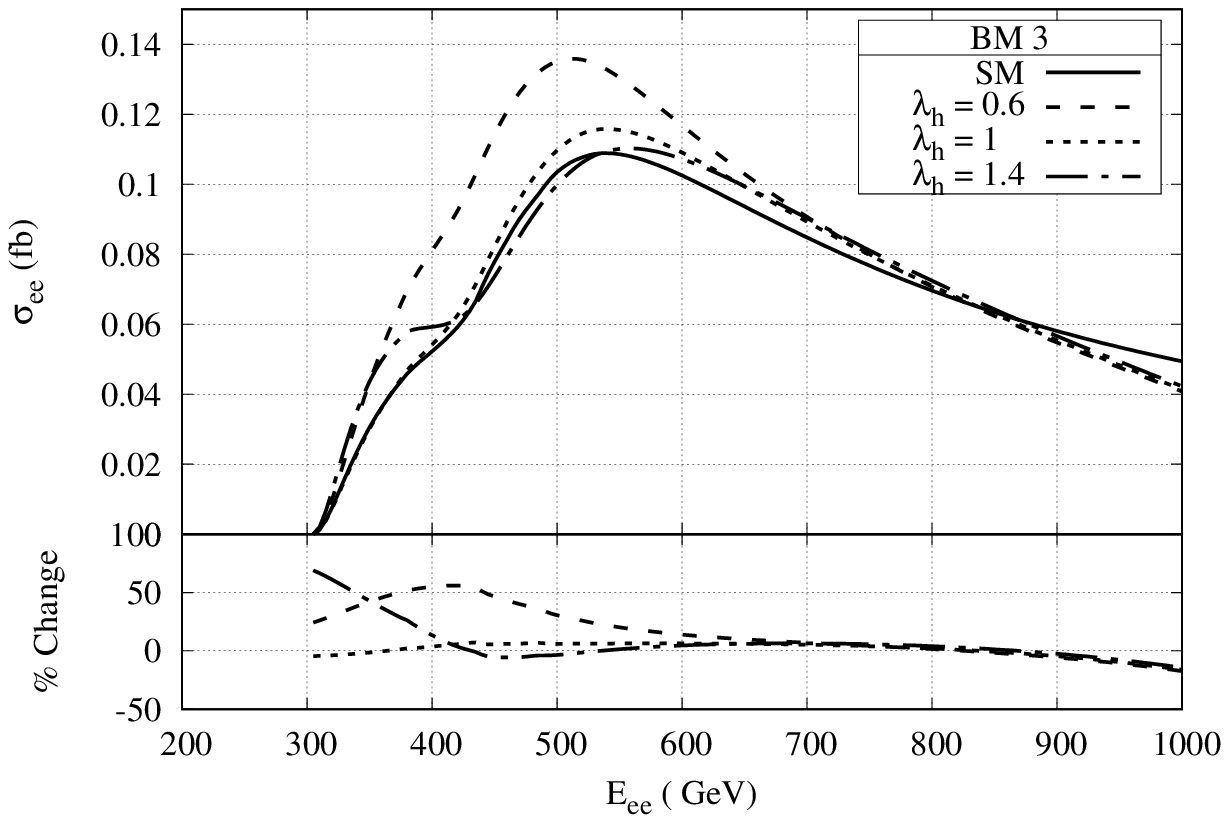,width=0.52\textwidth} 	
	\caption{$e^+ e^- \to h h$ cross-section in the model with an additional heavy scalar $H$. The benchmark 
	points (BP) for the above results are given in Table \ref{table:BM12}.}   	\label{fig:sec4_2} 
	\end{center}
\end{figure}

In Fig.~\ref{fig:sec4_2}, we show the $e^{+}e^{-}$ cross-sections for the three benchmark points, again for three values of the SM-Higgs quartic coupling, $\lambda_{H}=0.6$, 1, 1.4.
These benchmark points (defined in Table \ref{table:BM12}) are consistent with the present experimental constraints. 
The plots clearly show a sizeable enhancement around the mass of $H$, which also depends on the width of this state $\Gamma_{H}$.
The $\lambda_{H}$ dependence is only visible for low center of mass energies, below which the effect of the new diagrams become dominant.
For BM1 and BM2, which feature a lighter $H$, sizeable enhancements are expected at low energies, where the cross section can even double with respect to the SM.
In the BM3 case, the new scalar is too heavy to affect the total rate, while a significant enhancement of the cross section is still possible due to $\lambda_{H}$ and the large value of the new $\bar{t}thh$ quartic coupling, in a similar fashion in the case of minimal CHMs.

\section{The statistical sensitivity}\label{section:3}

We study the statistical sensitivity to the $\gamma\gamma\to hh$ process for wide regions of the collider energies at the PLC in the MCHM4 and MCHM5.
The statistical sensitivity $S_{stat}$ is defined by
\beq
S_{stat} = \frac{\left|N-N_{SM}\right|}{\sqrt{N_{obs}}} = \frac{L|\eta\sigma - \eta\sigma_{SM}|}{\sqrt{L(\eta\sigma+\eta_{BG}\sigma_{BG})}} \,,
\eeq
where $N$ and $N_{SM}$ are the expected number of events for the $\gamma\gamma \to hh$ process in the CHMs and SM respectively, and $N_{obs}$ is the observed total number of events in the CHMs including the back ground processes.
$\sigma$ and $\sigma_{SM}$ are the cross section of the Higgs boson production in the CHMs and SM, while $L$, $\eta$, $\eta_{BG}$, and $\sigma_{BG}$ are the integrated luminosity, the detection efficiency for the signal, the detection efficiency for backgrounds, and the cross section of background processes, respectively.

\begin{figure}[htb]
	\begin{center}
	\hspace*{-1cm}
		\epsfig{file=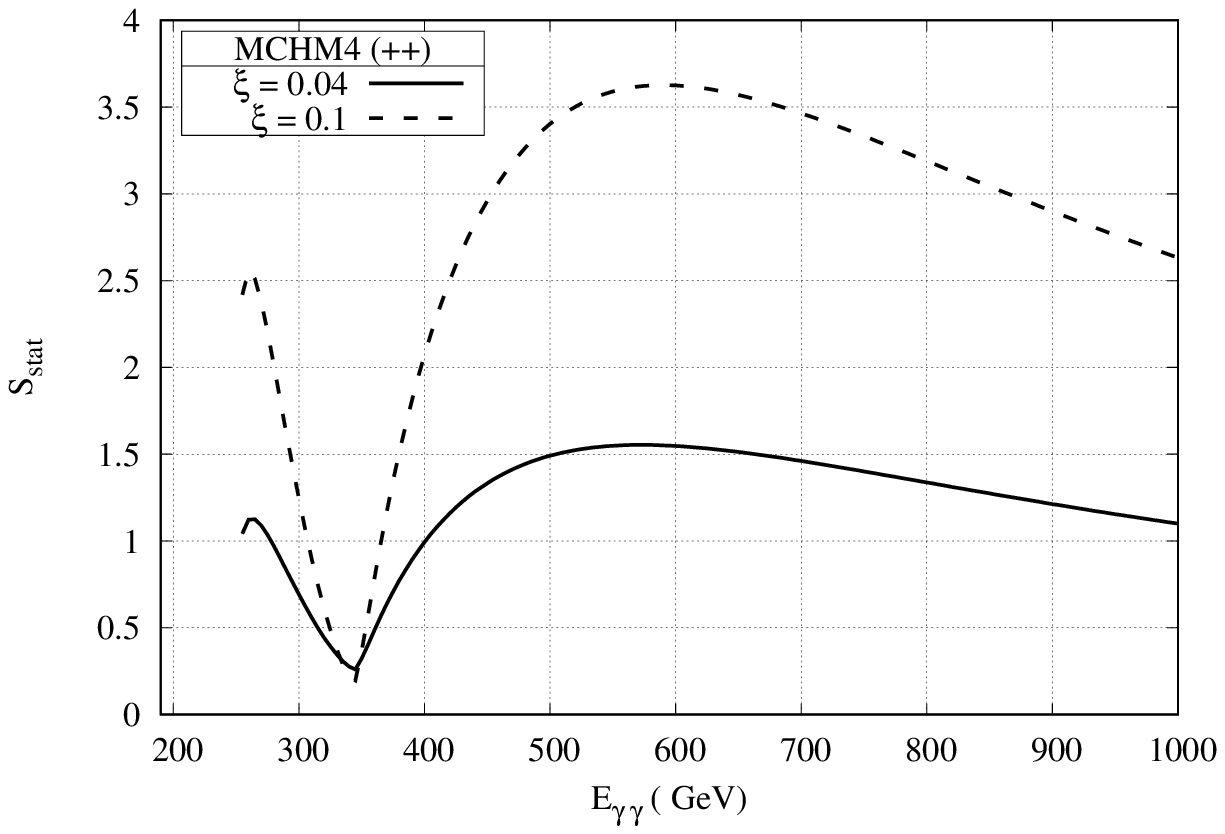,width=0.52\textwidth}
		\epsfig{file=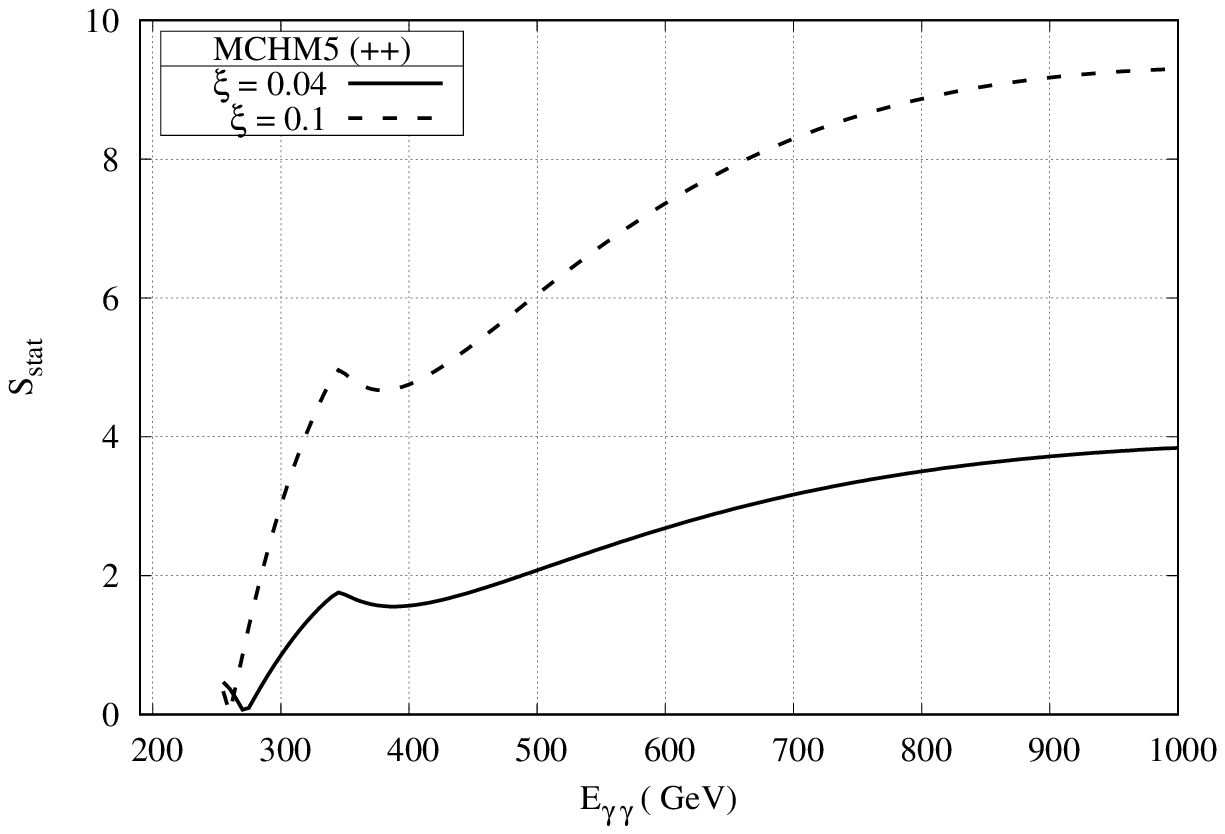,width=0.52\textwidth}
	\caption{Statistical Sensitivity ($S_{stat}$) in the MCHM4 (left panel) and MCHM5 (right panel) as a function of $E_{\gamma\gamma}$ energy for two values of $\xi=0.1, 0.04$.
	}   	\label{fig:sec3_1} 
	\end{center}
\end{figure}

In Fig.~\ref{fig:sec3_1}, we present the statistical sensitivity in the MCHM4 and MCHM5 at the PLC for two values of $\xi=0.1, 0.04$.
We assume that the efficiency of the particle tagging is 100\% and the BG processes are neglected ($\eta=1$ and $\eta_{BG}=0$) with an integrated luminosity $L=1$ ${\rm ab}^{-1}$, and $E_{\gamma\gamma}$ is the center of mass energy of the $\gamma\gamma$ system.
The plots show that the statistical sensitivity can be quite sizeable in particular the case of MCHM5 when the collision energy is limited to be lower than 1 TeV. In Ref.~\cite{Kawada:2012uy}, they point out that the $\gamma\gamma\to hh$ process in the SM with anomalous Higgs self-coupling can be observed with a statistical significance of about 5 $\sigma$ at the PLC at $L=2$ ${\rm ab}^{-1}$ against the large background processes such as $\gamma\gamma\to W^{+}W^{-}$, $ZZ$ and $b\bar{b}b\bar{b}$.
Therefore, the $\gamma\gamma\to hh$ process at the PLC would be expected useful to search the New Physics effects in CHMs as well as the anomalous Higgs self-coupling even though we consider these large back ground processes.

\section{Conclusions}\label{section:4}

In this work, we have studied the Higgs pair production process $\gamma\gamma\to hh$ in composite Higgs models, MCHM4, MCHM5 and CHMs with the heavy scalar, at the PLC.
We focused on composite models as they can provide novel Higgs pair production mechanisms and interference effects.
Our analysis show that these models can alter the SM prediction substantially, as photon collisions are sensitive to all modified Higgs couplings and 
effects stemming from the new quartic Higgs-fermion vertices.
The coupling responsible for these (quartic) vertices were absent in the SM and arise in CHMs due to the non-linear nature of the composite Higgs and 
 is one of the major reasons of enhancements in the photons initiated Higgs pair production.
On the other hand, modifications to the SM-like couplings do not depend much on the specific model, and only result in milder changes in the 
cross-sections.
In the MCHM5 benchmark with a large value of the Higgs-top quartic coupling, more than 20-30\% large enhancement of the $e^{+}e^{-} \to hh$ cross section is possible.
The presence of an additional scalar resonance $H$ opens up a new s-channel diagram,  affecting the same-sign helicity photon cross sections.
This can result in large enhancement of the $e^{+}e^{-} \to hh$ cross section by orders of magnitude for BM1 where $m_{H}=610$ GeV and $\Gamma_{H}=498$ GeV (large width).
The Higgs pair production process in $\gamma \gamma$ collision, therefore, is a key element for the discovery of deviations from the SM value.
This would be a strong indication of the composite structure underlying the Higgs sector.
In particular our results of the statistical sensitivity at the PLC in the composite models shows that they can be revealed at future lepton colliders, 
even if the energy is not sufficient to produce new resonances.

\subsection*{Acknowledgements}

AB, GC, AD, NG, FM and KS would like to thank CEFIPRA for the financial support on the project 
entitled  "Composite Models at the Interface of Theory and Phenomenology" (Project No. 5904-C). 
This work is also supported in part by the TYL-FJPPL program.


\end{document}